\documentclass[]{bmcart}

\usepackage{amsthm,amsmath}
\usepackage{bm}
\usepackage[utf8]{inputenc} 
\usepackage{graphicx}
\graphicspath{{./Figs/}}
\usepackage{cancel}
\usepackage{booktabs}

\startlocaldefs

\usepackage{upgreek}
\usepackage{graphicx}
\usepackage{subfigure}
\usepackage{color}

\usepackage{color}
\usepackage[normalem]{ulem} 
\newcommand{\Replace}[2]{\bgroup\noindent\textcolor{red}{\xout{#1} #2}\egroup\ignorespacesafterend}
\newcommand{\Delete} [1]{\bgroup\noindent\textcolor{red}{\xout{#1}}\egroup\ignorespacesafterend}
\newcommand{\Insert} [1]{\bgroup\noindent\textcolor{}{#1}\egroup\ignorespacesafterend}
\newcommand{\Comment}[1]{\definecolor{Mygray}{gray}{0.50}\bgroup\color{Mygray}\noindent#1\egroup\ignorespacesafterend}
\newcommand \Michael [1]{\bgroup\noindent[\textcolor{blue}{\textbf{Michael}: #1}]\egroup\ignorespacesafterend}
\newcommand \Stefan  [1]{\bgroup\noindent[\textcolor{blue}{\textbf{Stefan}: #1}]\egroup\ignorespacesafterend}

\DeclareMathAlphabet{\Ibb}{U}{msb}{m}{n}

 \newcommand{\BM}{{\boldsymbol{\mathnormal M}}}

 \newcommand{\BI}{{\boldsymbol{\mathnormal I}}}
 \newcommand{\Bomega  }{\ensuremath{\boldsymbol\omega}}

\newcommand{\taue}{\tau_{\rm ext}}

\newcommand{\Bb}{{\boldsymbol{\mathnormal b}}}

\newcommand{\Be}{{\boldsymbol{\mathnormal e}}}

\newcommand{\Bk}{{\boldsymbol{\mathnormal k}}}

\newcommand{\Bn}{{\boldsymbol{\mathnormal n}}}

\newcommand{\Bq}{{\boldsymbol{\mathnormal q}}}
\newcommand{\Br}{{\boldsymbol{\mathnormal r}}}

\newcommand{\Bv}{{\boldsymbol{\mathnormal v}}}

\newcommand{\Bbeta  }{\ensuremath{\boldsymbol\beta}}

\newcommand{\Beps    }{\ensuremath{\boldsymbol\epsilon}}

\newcommand \MZ [1] {\bgroup\noindent[\textcolor{blue}{\textbf{MZ}: #1}]\egroup\ignorespacesafterend}

\newcommand \Mdel [1] {\bgroup\noindent[\textcolor{red}{\textbf{Mdel}: #1}]\egroup\ignorespacesafterend}
\endlocaldefs

\newcommand \Madd [1] {\bgroup\noindent[\textcolor{blue}{\textbf{Madd}: #1}]\egroup\ignorespacesafterend}
\endlocaldefs

\begin{document}

\begin{frontmatter}

\begin{fmbox}
\dochead{Research}


\title{Cell structure formation in a two-dimensional density-based dislocation dynamics model}


\author[
   addressref={aff2,aff1},                   
]{\inits{RW}\fnm{Ronghai} \snm{Wu}}
\author[
   addressref={aff1},
   corref={aff1},                       
   email={michael.Zaiser@fau.de}
]{\inits{MZ}\fnm{Michael} \snm{Zaiser}}


\address[id=aff1]{
  \orgname{Department of Materials Simulation, Friedrich-Alexander Universität Erlangen-Nürnberg}, 
  \street{Dr.-Mack-str. 77},                     %
  \postcode{90762}                                
  \city{Fürth},                              
  \cny{Germany}                                    
}
\address[id=aff2]{
  \orgname{School of Mechanics, Civil Engineering and Architecture, Northwestern Polytechnical University}
  \postcode{710129}                                
  \city{Xian},                              
  \cny{P.R. China}                                    
}


\begin{artnotes}
\end{artnotes}

\end{fmbox}


\begin{abstractbox}

\begin{abstract} 
Cellular patterns formed by self-organization of dislocations are a most conspicuous feature of dislocation microstructure evolution during plastic deformation. To elucidate the physical mechanisms underlying dislocation cell structure formation, we use a minimal model for the evolution of dislocation densities under load. By considering only two slip systems in a plane strain setting, we arrive at a model which is amenable to analytical stability analysis and numerical simulation. We use this model to establish analytical stability criteria for cell structures to emerge, to investigate the dynamics of the patterning process and establish the mechanism of pattern wavelength selection. 
This analysis demonstrates an intimate relationship between hardening and cell structure formation, which appears as an almost inevitable corollary to dislocation dominated strain hardening. Specific mechanisms such as cross slip, by contrast, turn out to be incidental to the formation of cellular patterns. 
\end{abstract}




\end{abstractbox}
%

\end{frontmatter}



\section[Introduction]{Introduction}

Plastic deformation by dislocation motion is generally associated with dislocation patterning, leading to formation of heterogeneous dislocation arrangements. If multiple slip systems are active, dislocations form cellular structures where dislocation depleted 'cell interiors' are surrounded by dislocation rich 'cell walls' \cite{Szekely2002_MSEA}. Such cell structures show an almost universal scaling behavior (`law of similitude') which is independent of loading condition, material or temperature: the characteristic pattern wavelength $\lambda$ is proportional to the mean dislocation spacing (mds) $\rho_0^{-1/2}$ where $\rho_0$ is the spatially averaged dislocation density, and inversely proportional to the applied stress \cite{Rudolph2005_CRT, Sauzay2011_PMS}: $\lambda \propto \rho^{-1/2} \propto 1/\taue$. This behavior results directly from fundamental scaling invariance properties of dislocation systems as discussed by Zaiser and Sandfeld \cite{Zaiser2014_MSMSE}. Recent investigations \cite{Oudriss2016_IJP} indicate an even stronger form of the similitude principle according to which the components (cell walls, cell interiors) of cell structures obey the similitude principle separately, such that the wall thickness $\lambda_{\rm w}$ is related to the wall dislocation density by $\lambda_{\rm w} = C \rho_{\rm w}^{-1/2}$ and the cell dislocation density to the cell size $\lambda_{\rm c} = C \rho_{\rm w}^{-1/2}$, in such a manner that the proportionality coefficients $C$ are identical. We note in passing that, under very specific conditions which may be the exception rather than the rule (namely, deformation of fcc crystals with the loading axis oriented along a [100] direction), fractal cell patterns with a wide spectrum of length scales may emerge \cite{H"ahner1998_PRL}. However, even in these exceptional cases, the length scales defined by the upper and lower boundaries of the fractal scaling regime of cell sizes obey the "law of similitude" \cite{Zaiser1998_MSEA,H"ahner1998_MSEA}. 

Numerous models have been proposed for dislocation cell structure formation. Early models often relied on phenomenological similarities between dislocation patterns and other patterning phenomena, and used these analogies as a motivation to adopt equations drawn from other realms of science (e.g. spinodal decomposition \cite{Holt1970_JAP} and chemical patterning as described by reaction-diffusion models \cite{Walgraef1985_JAP}). These equations were adapted to dislocations in a manner that, seen with malevolent eyes, might be envisaged as a mere re-labeling exercise. It is not easy to see how, if at all, such models account for the specifics of dislocation topology, dislocation motion and dislocation interactions - for instance, it is immediately evident that the fundamental mode of dislocation motion under stress is not diffusion but directed glide. In recent years, efforts have been made to match chemical patterning inspired models more closely to actual dislocation processes, by distinguishing slip systems \cite{Pontes2006_IJP} and providing physically motivated reaction terms \cite{Aoyagi2013_IJP}. However, in all these models the problem remains that diffusion terms do not appropriately describe the glide of dislocations, which needs to be described by transport terms that are of a hydrodynamic rather than of a diffusion-like character, with important consequences to the nature of the emergent instabilities.

Discrete dislocation dynamics (DDD) simulation provides a powerful alternative to phenomenological ad-hoc models. DDD simulations faithfully represent the kinematics and interactions of dislocations and should be well suited for modelling dislocation pattern formation. While existing simulations \cite{Madec2002_SM,Hussein2016_JMPS} indicate that simulations of systems sufficiently large to allow for a quantitative investigation of pattern morphology alongside a reliable determination of pattern wavelengths may still be challenging, such limitations will be overcome with time simply due to the expected increase in available computing power. 

However, from an epistemological point of view the ability to provide a more or less faithful {\em in vitro} simulation of a real process should not be confounded with understanding: a sufficiently complex simulation may encompass, besides essential, a large amount of redundant features and it may not be easy to decide which features of the collective dynamics are at the core of a collective phenomenon such as dislocation cell structure formation, and which are incidental to it. Rather than pursuing accuracy in detail, our own modelling strategy therefore is heavily poised towards simplicity -- while at the same time we make sure that the most essential kinematic features and the structure of the interactions are represented correctly. Mathematical simplicity of the model allows us to obtain some results in an analytical or semi-analytical manner, and renders the essential features of the dynamics more transparent. To this end we rely on a most basic version of density based dislocation dynamics in multiple-slip conditions. We start from the model used by Zaiser, Groma and co-workers \cite{Groma2016_PRB,Wu2017_PRB} for analysing the conditions for pattern formation in single slip, and generalize this to symmetrical double slip along lines proposed in earlier work of Groma and co-workers \cite{Yefimov2005_IJSS,Limkumnerd2008_PRB}. This framework not only provides us with some degree of analytical tractability but also with a solid theoretical foundation: The equations we use have been rigorously derived from statistical averaging of the underlying discrete dynamics \cite{Groma2003_AM, Valdenaire2016_PRB} and can be related via variational calculus to the statistically averaged energy functional of the dislocation system \cite{Zaiser2015_PRB,Groma2016_PRB}. Moreover, predictions obtained with these equations for size-dependent deformation in small samples and/or constrained geometries have been shown to be in quantitative agreement with discrete dislocation dynamics simulations \cite{Yefimov2004_JMPS,Yefimov2005_IJSS}. This makes us confident that the mathematical framework we used indeed captures essential features of dislocation dynamics under load.

We note that other, more complex versions of density-based continuum dislocation dynamics have been applied to the patterning problem. Some of these approaches consider geometrically necessary dislocations only \cite{Limkumnerd2008_JMPS,Chen2013_IJP}. However, during the early stages of deformation the dislocations in the cell walls have near-zero net Burgers vector: they are predominantly {\em not} geometrically necessary dislocations. Application of such models to early stages of cell structure formation is therefore possible only if the spatial resolution is well below the actual dislocation spacing such that Burgers vectors do not cancel out. If one makes this numerical effort the results can be impressive \cite{Xia2015_MSMSE} and capture dislocation processes in three-dimensional dislocation patterns in detail \cite{lin2020implementation}. A more coarse grained model that allows for co-existence of dislocations of different Burgers vector in the elementary volume but nevertheless captures effects of three-dimensional curvature was proposed by Sandfeld and Zaiser \cite{Sandfeld2015_MSMSE}. An interesting work was recently published by Grilli et al. \cite{Grilli2018_IJP}. These authors consider two models which allow for dislocations of different Burgers vector in the same elementary volume, which are described by a set of densities obeying transport equations and applied to labyrinth-like patterns emerging under cyclic loading. These works are conceputally more complex than the present one, as they consider three-dimensionally curved dislocations \cite{Sandfeld2015_MSMSE}, distinguish various orientations \cite{Grilli2018_IJP}, and include essentially three-dimensional processes such as junction formation \cite{Grilli2018_IJP} and cross slip \cite{Xia2015_MSMSE}. While these approaches are interesting in their own right, we demonstrate in the present paper that the added complexity is actually not essential for cell structure formation or dislocation patterning as such. In the following we first briefly introduce the governing equations of our model and then provide a stability analysis that allows us to establish necessary conditions for cell pattern formation. We show the results of numerical simulations of the evolution equations and compare our findings to experimental data. Finally we provide a conclusion where we discuss implications of our findings in view of some commonly held ideas regarding the nature of dislocation patterns and the requirements for their formation.

\section{Model Equations}
We consider a crystal deforming in plane strain where two orthogonal slip systems are active. System 1 has Burgers vector $\Bb_1 = b\Be_x$ and slip plane normal $\Bn_1 = \Be_y$, and system 2 has Burgers vector $\Bb_2 = b \Be_y$ and slip plane normal $\Bn_2 = \Be_x$. The shear strains on the two slip systems are denoted as $\gamma_1$ and $\gamma_2$. The plastic distortion is then given by 
\begin{flalign}
\label{betapl}
&\Bbeta^{\rm pl} = \gamma_1 [\Be_y\otimes\Be_x] + \gamma_2  [\Be_x\otimes\Be_y].
\end{flalign}
We define the plastic strain $\Beps^{\rm pl}$ and plastic rotation $\Bomega^{\rm pl}$ as the symmetric and anti-symmetric parts of $\Bbeta^{\rm pl}$. 
These are given by 
\begin{flalign}
\label{epsomega}
&\Beps^{\rm pl} = \frac{\gamma}{2}  [\Be_y\otimes\Be_x + \Be_x\otimes\Be_y],\\
&\Bomega^{\rm pl} = \frac{\omega}{2}  [\Be_y\otimes\Be_x - \Be_x\otimes\Be_y].
\end{flalign}
where $\gamma = \gamma_1 + \gamma_2$ and $\omega = \gamma_1 - \gamma_2$. 

Both slip systems contain straight parallel edge dislocations gliding in the directions of the respective Burgers vectors. We assume that each system contains equal numbers of positive and negative dislocations with the corresponding dislocation densities denoted as $\rho^{\pm}_{1/2}$ where the upper label distinguishes positive and negative dislocations, and the lower label distinguishes the two slip systems. Positive dislocations move under the action of a positive resolved shear stress in the positive Burgers vector directions, and negative dislocations move under the same shear stress in the negative Burgers vector directions, $\Bv^{\pm}_{1/2} = \pm v^{\pm}_{1/2} \Bb_{1/2}/b$ where $v^{\pm}_{1/2}$ are scalar velocities.

In the spirit of defining a minimal model, we neglect dislocation reactions (which anyway, for energetic reasons, are not expected to yield stable products), dislocation multiplication and annihilation. The dislocation densities are thus conserved quantities which obey the continuity equations
\begin{flalign}
&\frac{\partial \rho_1^+}{\partial t} = - \partial_x (\rho_1^+ v_1^+),\nonumber\\
&\frac{\partial \rho_1^-}{\partial t} =   \partial_x (\rho_1^- v_1^-),\nonumber\\
&\frac{\partial \rho_2^+}{\partial t} = - \partial_y (\rho_2^+ v_2^+),\nonumber\\
&\frac{\partial \rho_2^-}{\partial t} =   \partial_x (\rho_2^- v_2^-).
\label{eq:transport}
\end{flalign}
The dislocation velocities for these four types of dislocations are assumed to be linearly proportional to respective, effective shear stresses 
${\cal T}_{i}^{s}$ where the index $i \in \{1,2\}$ distinguishes the two slip systems and $s \in \{-1,1\}$ distinguishes the two signs of the dislocations:
\begin{flalign}
&v^{s}_{i}(\vec{r},t) = \frac{b}{B} {\cal T}_i^s(\vec{r},t).
\label{eq:velocities} 
\end{flalign}
In these equations, $B$ is the dislocation drag coefficient. A closed mathematical model is then specified by relating the effective shear stresses to the dislocation densities. In line with the single-slip model of Groma and co-workers \cite{Groma2016_PRB,Wu2017_PRB}, we consider the effective driving stresses ${\cal T}_i^s$ to result from the combination of sign-dependent local driving stresses $\tau_{i}^{s,\rm dr}$ and friction stresses $\tau_{i}^{s,\rm f}$: 
\begin{equation}
{\cal T}_i^s = \left\{\begin{array}{l}
 \rm{sign}(\tau_{i}^{s, \rm dr}) \left(|\tau_{i}^{s,\rm dr}| - \tau_{i}^{s,\rm f}\right) \quad{\rm if}\quad |\tau_{i}^{s,\rm dr}| - \tau_{i}^{s,\rm f} > 0\\
0 \quad {\rm otherwise}
\end{array}\right.
\label{eq:taueff}
\end{equation}
The driving stresses combine the resolved shear stress $\tau_i$  in the respective slip system with corrections describing short-range dislocation interactions associated with the mutual arrangement of individual dislocations (dislocation correlations) according to
\begin{flalign}
&\tau_{i,s}^{\rm dr} = \tau_{i} + \tau_{i}^{\rm b} + s\tau_{i}^{\rm d}.
\label{eq:taudrive}
\end{flalign}
We discuss the three stress contributions in this equation separately:
\begin{enumerate}
\item
The resolved shear stress $\tau_{i}$ arises from the superposition of stresses caused by external tractions and internal stresses associated with the plastic eigenstrains -- in other words, it is found by solving a standard elastic-plastic problem. The considered slip geometry has the peculiarity that this stress is the same in both slip systems and equals the $xy$ component of the stress tensor, $\tau_{1} = \tau_2 = \sigma_{xy}$. In our calculations, we consider a bulk system with periodic boundary conditions and calculate this stress from the plastic strain $\gamma$ using a Green's function formalism \cite{Zaiser2005_JSTAT,Wu2017_PRB}:
\begin{flalign}
&\tau(\Br) = \tau_{\rm ext} - \int \gamma(\Br') {\cal G}(\Br-\Br') {\rm d}^2 r' \label{eq:intstress}
\end{flalign}
where $\tau_{\rm ext}$ is a spatially constant external stress arising from remote tractions acting on the infinite contour, and ${\cal G}$ is an interaction kernel function with the Fourier transform
\begin{flalign}
&{\cal G}(\Bk) = \frac{G}{\pi(1-\nu)} \frac{k_x^2k_y^2}{k^4} = G T(\Bk). \label{eq:Gk}
\end{flalign}
$G$ is the shear modulus of the material, $\nu$ is Poisson's ratio, and $k_x$ and $k_y$ are components of the Fourier wave-vector with modulus $k$. 
\item
The 'back stresses' $\tau_i^{\rm b}$ stem from the mutual correlation of dislocations of the same sign and counter-act their accumulation. For single slip on some slip system $i$ the back stress is given by  
\begin{flalign}
&\tau_i^{\rm b}(\Br) = -G \frac{D}{\rho_i} (\Bb_i.\nabla) \kappa_i(\Br), \label{eq:backstress}
\end{flalign}
where $D$ is a non-dimensional factor of the order of unity and $\rho = \rho^+ + \rho^-$ is the total dislocation density on the considered slip system. The local excess density $\kappa_i$ is given by the difference of positive and negative dislocation densities and relates to the slip gradient on the slip system $i$ via
\begin{flalign}
&\kappa_i = \rho_i^+ - \rho_i^- = - \frac{1}{b^2} \Bb_i.\nabla \gamma_i.
\end{flalign}
For multiple slip situations as considered here, Linkumnerd et al \cite{Limkumnerd2008_PRB} use a statistical-mechanical model of the density cross correlation functions to derive instead of Eq. \ref{eq:backstress} the superposition relations  
\begin{flalign}
&\tau_i^{\rm b}(\Br) = - GD \sum_j \frac{\cos \theta_{ij}}{\rho_j} (\Bb_i.\nabla) \kappa_i(\Br), \label{eq:backstressmod}
\end{flalign}
where $\theta_{ij}$ are the angles between the Burgers vectors (slip directions) of slip system pairs. For the geometry considered here, $\cos \theta_{ij} = \delta_{ij}$  and hence, Eqs. (\ref{eq:backstress}) and (\ref{eq:backstressmod}) are equivalent. 
\item
Accordingly, we consider the 'diffusion stresses' $\tau_i^{\rm d}$ to be given by
\begin{flalign}
&\tau_i^{\rm d}(\Br) = - GA \frac{1}{\rho_i}(\Bb_i. \nabla)\rho_i(\Br),  \label{eq:diffstress}
\end{flalign}
where $A$ is another nondimensional factor of the order of unity. The terminology 'diffusion stresses' is used because this stress, if inserted via Eqs (\ref{eq:taudrive}), (\ref{eq:taueff}), (\ref{eq:velocities}) into the transport equations Eq. (\ref{eq:transport}), gives rise to diffusion-like contributions to the evolution of the total dislocation densities $\rho_i$. 
\end{enumerate}
All three stress contributions can be derived from the energy functional of the dislocation system, as discussed in detail by Groma et al 
\cite{Groma2016_PRB}, hence, they are associated with stored energy contributions. 

It remains to specify the friction-like stresses $\tau_{i}^{s,\rm f}$. In generalization of the expression derived by Grpma et al \cite{Groma2016_PRB} for single slip, we assume these stresses in the form
\begin{flalign}
&\tau^{s,\rm f}_{i}  =\alpha G b \sqrt{\sum_j H_{ij} \rho_j} \left(1 - s \frac{\kappa_i}{\rho_i}\right) , 
\label{eq:flowstress}
\end{flalign}
where the latent hardening matrix $H_{ij}$ describes slip system interactions. The dependency on the $\kappa_i$ accounts for the fact that excess dislocations cannot be pinned by dislocations of the same slip system (the net force on the excess cannot become zero). For more details see Ref. \cite{Wu2017_PRB}. 

For the present system, the resolved shear stresses induced by a dislocation in both slip systems are equal, hence, it is reasonable to set $H_{ii} = H_{ij} = 1$ leading to 
\begin{flalign}
&\tau^{\rm f}_{i,s}    =\alpha G b \sqrt{\rho} \left(1- s \frac{\kappa_i}{2\rho_i}\right) 
\label{eq:flowstresssimple}
\end{flalign}
where $\rho = \rho_1 + \rho_2$ is the total dislocation density. The friction stresses are of a different nature from the driving stresses: they represent friction-like stresses that are associated with dissipated, not with stored energy contributions. While these stresses arise naturally from direct averaging of the dislocation interactions, they cannot be derived from an energy functional but need to be added 'by hand' to an energy-based formalism where they enter in terms of a non-trivial, nonlinear mobility function with a mobility threshold \cite{Groma2016_PRB}. The functional form of these stresses is that of Taylor stresses; in physical terms, they represent the mutual trapping of dislocations into dipolar or multipolar configurations. Their dependency on the $\kappa_i$ reflects the fact that the presence of an excess of dislocations of one sign implies reduced pinning of the majority and enhanced pinning of the minority population. 

Assembling all stress contributions, we find that the four dislocation density species under consideration fulfill, under the assumption that the local effective stress is positive and the system is everywhere in the flowing phase, the respective continuity equations
\begin{flalign}
\frac{\partial\rho_{i,s}}{\partial t} &= - \frac{1}{B} s \Bb_i. \nabla \left\{\rho_{i,s} \left[\tau_{\rm ext} - \int \sum_i \gamma_i(\Br') {\cal G}(\Br-\Br') {\rm d}^2 r'\right.\right.\nonumber\\
&-
\left.\left.
\frac{G}{\rho_i} \Bb_i. \nabla \left[D\kappa_i + s A \rho_i\right] - \alpha G b \sqrt{\sum_{i,s} \rho_{i,s}}\left(1 - s \frac{\kappa_i}{\rho_i}\right)\right]\right\}. 
\label{eq:densityevolution}
\end{flalign}
The strains $\gamma_i$ evolve according to
\begin{flalign}
\frac{\partial\gamma_{i}}{\partial t} &= \frac{b^2}{B} \sum_{s} \left\{\rho_{i,s} \left[\tau_{\rm ext} - \int \sum_i \gamma_i(\Br') {\cal G}(\Br-\Br') {\rm d}^2 r'\right.\right.\nonumber\\
&-
\left.\left.
\frac{G}{\rho_i} \Bb_i. \nabla \left[D\kappa_i + s A \rho_i\right] - \alpha G b \sqrt{\sum_{i,s} \rho_{i,s}}\left(1 - s \frac{\kappa_i}{\rho_i}\right)\right]\right\}.
\label{eq:strainevolution}
\end{flalign}
Before we proceed to analyze the model equations, it is important to comment on the nature and meaning of the non-dimensional parameters $A,D$, and $\alpha$ which enter the model in addition to the physical constants $G,b,\nu$, and the drag coefficient $B$. All three parameters $A,D$, and $\alpha$ characterize correlations in the positions of individual dislocations and can in principle be evaluated in terms of integrals over dislocation-dislocation correlation functions, see their derivations in Refs. \cite{Groma2003_AM, Limkumnerd2008_PRB, Groma2016_PRB, Valdenaire2016_PRB}. All these parameters are of the same order of magnitude as they characterize the arrangement of close dislocations whose positions, owing to their mutual interactions, are strongly correlated. Specifically, $\alpha$ is proportional to the characteristic spacing of dislocations that have trapped each other into dipolar or multipolar configurations, measured in units of the typical spacing of dislocations of the same slip system in the surrounding of a given spatial point -- of course, as such $\alpha$ is nothing but the well known Taylor factor.  If the dislocation arrangement is thought of as an assembly of isolated dipoles of height $h$, then $\alpha = (8 \pi (1-\nu) (h\sqrt{\rho})$, but in more general circumstances, this factor needs to be modified to account for the influence of dislocations surrounding the dipole. The parameters $A$ and $D$ have an analogous interpretation, but 'probe' different aspects of short-range interactions: While $\alpha$ mainly captures the trapping effect of dipole-like interactions, $D$ characterizes the interactions between dislocations of the same sign in piled-up configurations, which cause a net stress if there is a gradient in the 'geometrically necessary' density $\kappa$. Finally, $A$ which controls the 'diffusion stress' accounts for the fact that dipoles and multipoles have finite extension, such that dislocation density cannot localize down to arbitrary narrow scales. In summary, all three factors are proportional to spacings of {\em individual} dislocations, with $\alpha$ mainly characterizing the spacing of slip planes of adjacent dislocations, $D$ spacing of dislocations of the same sign in piled up configurations, and $A$ the extension of dipoles and multipoles in glide direction. 

Understanding the physical nature of the constants $\alpha,D,A$ is also beneficial for the physical interpretation of the respective stress contributions. Breaking of dipoles and formation of new ones is a dissipative process that occurs as soon as the local stress exceeds the dipole breaking stress, hence, the associated stress contribution has friction-like characteristics. Piling up dislocations against an obstacle, by contrast, leads to storage of energy that can be recovered if the stress causing the pile up is removed or reversed, hence, the associated energy contribution enters an appropriately averaged internal energy functional. The same is true for the work expended in compressing or expanding dipolar and multipolar configurations. It is in line with these intuitive arguments that, upon formal statistical averaging of the elastic energy of a dislocation system \cite{Zaiser2015_PRB}, the resulting density based functional allows to recover through variational calculus both the 'back stress' and the 'diffusion stress' but not the 'friction stress' \cite{Groma2016_PRB}.
\section{Stability analysis}

\subsection{Reference state}

We consider pattern formation first in an analytical framework where we focus on infinitesimal perturbations of a spatially homogeneous reference state where $\rho_{i,s} = \rho_0/4 \;\forall \,\{i,s\}$ and $\gamma_i = \gamma_0/2 \;\forall \;i$. At this stage we envisage loading by a temporally constant applied stress $\taue$. Depending on the level of stress, two situations need to be distinguished:
(i) If $\taue < \alpha G b \sqrt{\rho_0}$ then all velocities in the reference state are zero, hence, $\gamma=0$ is constant in space and time and $\rho_{i,s}=\rho_0/4$ is a stationary solution of the evolution equations that is stable with respect to infinitesimal perturbations. (ii) If $\taue > \alpha G b \sqrt{\rho_0}$ we are in a flowing phase. In this case the dislocations move with homogeneous and stationary velocity $v_0 = (b/B) (\taue - \alpha G b \sqrt{\rho_0})$ and the slip system strains increase linearly in time, $\partial_t \gamma_i = \dot{\gamma}_0/2 = \rho_0 b v_0/2$. The stability of this flowing state is analyzed in the following. 

In our analysis we have a choice of variables. Instead of the four densities $\rho_{i,s}$ we may use the total and excess dislocation densities on the two slip systems, $\rho_i = \sum_s \rho_{i,s}$ and $\kappa_i = \sum_s s \rho_{i,s}$. Furthermore, instead of the excess dislocation densities we may alternatively consider the slip variables $\gamma_i$ which relate to the former via $\kappa_i = \Bb_i \nabla \gamma_i/b^2$. This is the choice we make, i.e., we consider the problem in terms of the four variables $\rho_i,\gamma_i$, $i \in \{1,2\}$. 

\subsection{Dimensionless scaling}
In the following we switch to a dimensionless formulation which helps to see the influence of all model constituents more easily. Only the final results are stated here, for detailed information and derivations see Refs. \cite{Zaiser2014_MSMSE, Sandfeld2015_MSMSE}.
We define the scaling relations between quantities with physical units and their dimensionless counterparts (indicated by a tilde) as $\tau  =C_\tau \tilde\tau$ (for stresses), $\rho^s  =C_\rho \tilde\rho^s$ (for dislocation densities), $x =C_x \tilde x$ (for lengths), and $\gamma = C_{\gamma} \tilde{\gamma}$, 
with the scaling factors  
\begin{flalign}\label{eq:scaling_factors1}
C_\tau= \alpha G b \sqrt{\rho_0}, \quad 
C_\rho=\rho_0, \quad 
C_x=\rho_0^{-1/2}, \quad
C_{\gamma} = b \rho_0^{1/2}.
\end{flalign}
Furthermore, we scale velocities in units of $C_v = b C_{\tau}/B$, which implies a scaling for time according to 
$t = C_t \tilde{t}$ with $C_t = C_x/C_v$. In non-dimensional form the equations of motion become 
\begin{flalign}
\frac{\partial\rho_{i,s}}{\partial t} &= - s\nabla_i \left\{\rho_{i,s} \left[\tilde{\tau}_{\rm ext} - \int \sum_i \gamma_i(\Br') \tilde{T}(\Br-\Br') {\rm d}^2 r'\right.\right.\nonumber\\
-&
\left.\left.
\frac{1}{\rho_i} \nabla_i \left[\tilde{D}\kappa_i + s \tilde{A} \rho_i\right] - \sqrt{\sum_{i,s} \rho_{i,s}}\left(1 + s \frac{\kappa_i}{\rho_i}\right)\right]\right\}. \\
\frac{\partial\gamma_{i}}{\partial t} &= \sum_{s} \left\{\rho_{i,s} \left[\tilde{\tau}_{\rm ext} - \int \sum_i \gamma_i(\Br') \tilde{T}(\Br-\Br') {\rm d}^2 r'\right.\right.\nonumber\\
-&
\left.\left.
\frac{1}{\rho_i}\nabla_i \left[\tilde{D}\kappa_i + s \tilde{A} \rho_i\right] - \sqrt{\sum_{i,s} \rho_{i,s}}\left(1 + s \frac{\kappa_i}{\rho_i}\right)\right\}\right).
\label{eq:scaledevolution}
\end{flalign}
where we have dropped the tildes on all variables and introduced the notations $\nabla_i = (\Bb_i.\nabla/b)$, $\tilde{D} = D/\alpha$, and $\tilde{A} = A/\alpha$. The scaled stress kernel is given by $\tilde{T} = T/(\alpha \rho_0)$ and has in scaled variables ($k \to k/\sqrt{\rho_0}$) the Fourier transform
\begin{flalign}
\tilde{T}(\Bk) = \frac{1}{\alpha\pi(1-\nu)} \frac{k_x^2k_y^2}{(k_x^2 + k_y^2)^2} = T_0 \frac{k_x^2k_y^2}{(k_x^2 + k_y^2)^2} . \label{eq:T}
\end{flalign}

\subsection{Linearized evolution equations}
We now write down the equations of evolution for small perturbations $\delta \rho_i, \delta \gamma_i$ of our reference state $\rho_{i,0}=1/2, \gamma_{i,0} = \gamma_0/2$. In linear approximation these perturbations are given by
\begin{flalign}
\frac{\partial\delta\rho_{i}}{\partial t} &= \nabla_i^2 (\tilde{A} \delta \rho_i + \taue \delta \gamma_i) ,\\
\frac{\partial\delta\gamma_{i}}{\partial t} &= (\taue - 1) \delta \rho_i - \frac{1}{4} \sum_j \delta \rho_j \nonumber\\
&+ \tilde{D} \nabla_i^2 \delta \gamma_i - \rho_i^0 \int \sum_j \delta \gamma_j(\Br') \tilde{T}(\Br-\Br') {\rm d}^2 r'
\label{eq:deltaevolution}
\end{flalign}
Defining the state vector $\delta \Bq = [\delta\rho_1,\delta \gamma_1, \delta \rho_2, \delta \gamma_2]$ and using the Fourier Ansatz
$\delta \Bq = \Bq(\Bk) \exp(i\Bk.\Br)$, we write these equations in matrix form:
\begin{flalign}
\frac{\partial}{\partial t}\Bq(\Bk) = \BM.\Bq(\Bk) 
\end{flalign}
with 
\begin{flalign}
\BM = \left[ \begin{array}{llll}
-\tilde{A} k_x^2 &  - \taue k_x^2 & 0 & 0\\
\taue - \frac{5}{4} & - \frac{\tilde{T}}{2} - \tilde{D} k_x^2 &  - \frac{1}{4} & - \frac{\tilde{T}}{2} \\
0 & 0 & -\tilde{A} k_y^2 &  - \taue k_y^2 \\
-1/2 & - \frac{\tilde{T}}{2}  & \taue - \frac{3}{2} & - \frac{\tilde{T}}{2} - \tilde{D} k_y^2 \\
\end{array} \right]
\end{flalign}
We now first investigate two simple cases where the eigenvalues can be computed analytically in a straightforward manner. 

\subsection{Symmetrical case}
We first study the eigenvectors and eigenvalues of this matrix for the symmetrical case $k_x = k_y = k/\sqrt{2}$. 
The matrix $\BM$ simplifies to 
\begin{flalign}
\BM = \left[ \begin{array}{llll}
- \frac{\tilde{A}}{2} k^2  & - \frac{\taue}{2} k^2 & 0 & 0\\
\taue - \frac{5}{4} &  - \frac{T_0}{8} - \frac{\tilde{D}}{2} k^2 &  - \frac{1}{4}  & - \frac{T_0}{8} \\
0 & 0 & -\frac{\tilde{A}}{2} k^2 &  - \frac{\taue}{2} k^2\\
-\frac{1}{4} & - \frac{T_0}{8} & \taue - \frac{5}{4} & -\frac{T_0}{8} - \frac{\tilde{D}}{2}k^2 \\
\end{array} \right]
\end{flalign}
The eigenvectors of this matrix have the structure $\Bq_1 = \pm \Bq_2$ where $\Bq_1 = [\delta \rho_1,\delta \gamma_1], \Bq_2=[\delta \rho_2,\delta \gamma_2]$. We 
first consider the "-" case. The matrix equation then reduces to $\BM^- .\Bq_1 = \Lambda^- \Bq_1$ where
\begin{flalign}
\BM^- = \left[ \begin{array}{ll}
-\frac{\tilde{A}}{2} k^2&  - \frac{\taue}{2} k^2\\
\taue - 1 & - \frac{\tilde{D}}{2} k^2\\
\end{array} \right]
\end{flalign}
The eigenvalues fulfil the characteristic equation
\begin{flalign}
(-\frac{\tilde{A}}{2} k^2 - \Lambda^-)(- \frac{\tilde{D}}{2} k^2 - \Lambda^-) + \frac{\taue(\taue-1)}{2} k^2 = 0
\end{flalign}
Since $\taue > 1$ in the flowing phase and both $\tilde{A}$ and $\tilde{D}$ are positive, both roots of this equation have negative real parts for all $k$ and $\taue$, 
hence, no instability can occur. In the "+" case we get $\BM^+ .\Bq_1 = \Lambda^+ \Bq_1$ where
\begin{flalign}
\BM^- = \left[ \begin{array}{ll}
-\tilde{A} k^2&  - \frac{\taue}{2} k^2\\
\taue - \frac{3}{2}&  - \frac{T_0}{8} -\tilde{D} k^2\\
\end{array} \right]
\end{flalign}
The eigenvalues then fulfil the characteristic equation
\begin{flalign}
(-\frac{\tilde{A}}{2} k^2 - \Lambda^-)(- \frac{\tilde{D}}{2} k^2 - \frac{T_0}{8}- \Lambda^+) + \frac{\taue}{2}(\taue-\frac{3}{2}) k^2 = 0
\end{flalign}
An unstable wavelength band may in that case occur if $1 \leq \taue$ and $8\taue(3/2-\taue) > T_0 \tilde{A}$. 
This band is comprised between the wavelengths $k=0$ and $k=k_{\rm c}^{[11]}$ where
\begin{flalign}
(k_{\rm c}^{[11]})^2 = \frac{8\taue(3/2-\taue) - T_0\tilde{A}}{4\tilde{A}\tilde{D}}.
\end{flalign}

\begin{figure}[htp]
	\centering
	\hbox{}
	\includegraphics[width=0.8\textwidth]{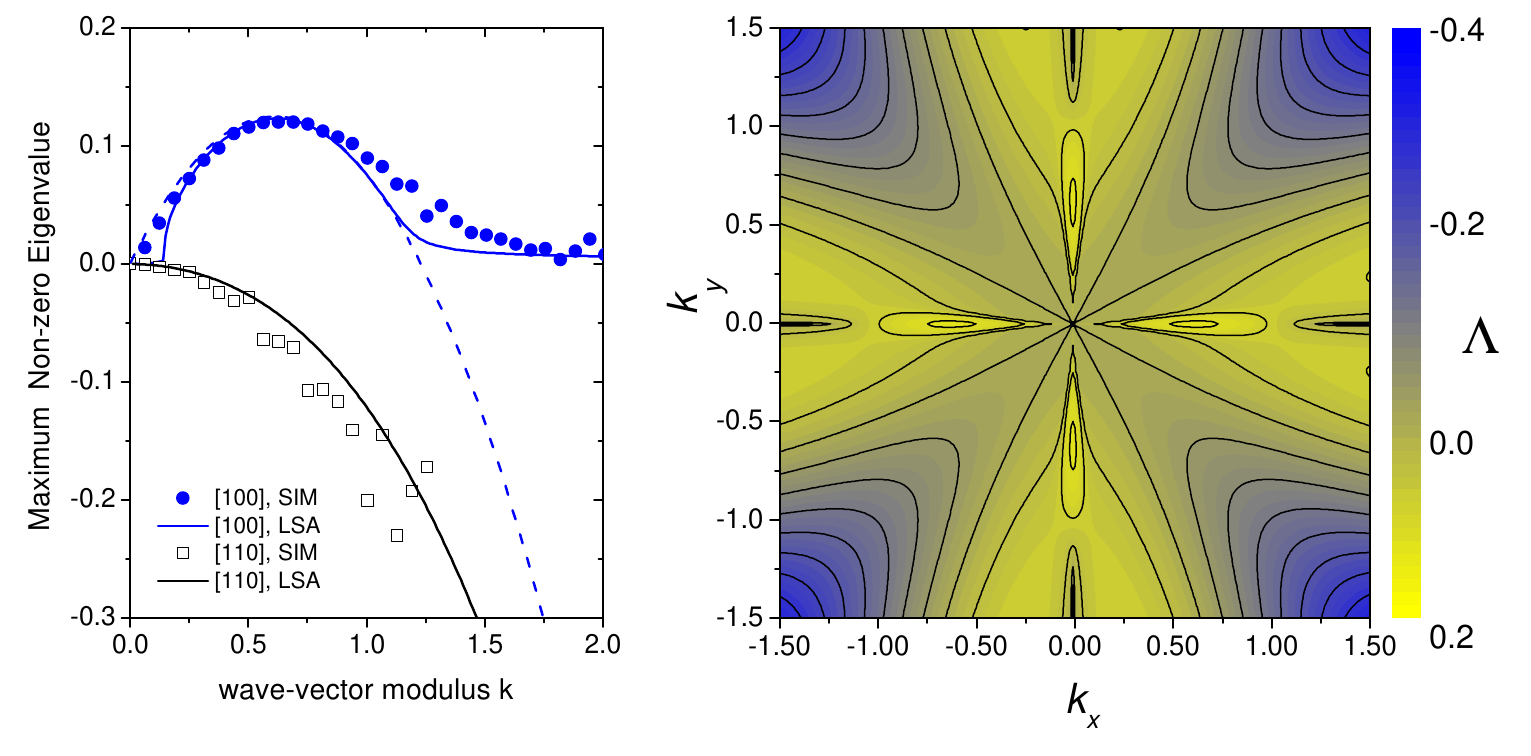}\hfill
		\caption{\label{fig:instability}
		Left: Growth rates of fluctuations; dashed blue line: growth rate for wave-vectors aligned with the [10] lattice directions according to linear stability analysis (LSA), 
		blue line: growth rate for fluctuations near [10], black line: growth rate for wave-vectors flaligned with the [11] lattice directions; discrete symbols: 
		growth  rates deduced from Fourier modes of the numerical solution for a Gaussian white noise as initial condition. Right: growth rates for fluctuations according to 
		LSA   over the entire domain of wave vectors. Parameters: $A=D=0.1$, $\alpha = 0.3$, $\taue = 1.1$; }
\end{figure}

\subsection{Fluctuations along the cube axis}

Next we consider the case where the fluctuation wave vectors are aligned with the $x$ axis, $k_x = k, k_y = 0$ (the opposite case is symmetry equivalent). 
The matrix $\BM$ simplifies to 
\begin{flalign}
\BM = \left[ \begin{array}{llll}
- \tilde{A} k^2  & - \taue k^2 & 0 & 0\\
\taue - \frac{5}{4} & - \tilde{D} k^2 &  -\frac{1}{4} & 0 \\
0 & 0 & 0 &  0 \\
-\frac{1}{4} & 0 & \taue - \frac{5}{4} & 0 \\
\end{array} \right]
\end{flalign}
The characteristic equation is obtained by setting the determinant of the matrix $\BM - \Lambda \BI$ to zero. Expanding the determinant with respect to the last column gives the straightforward result
\begin{flalign}
\Lambda^2 \left[(-\tilde{A} k^2 - \Lambda^-)(- \tilde{D} k^2 - \Lambda^-) + \taue(\taue-5/4) k^2\right] = 0
\end{flalign}
This characteristic equation is, but for the factor $\Lambda^2$ and the slightly different scaling, similar to the characteristic equation obtained for instabilities on a single slip system, hence, the results of Groma et al \cite{Groma2016_PRB,Wu2017_PRB} can be transferred. An unstable wavelength band occurs if $1 \leq \taue$ and $\taue < 5/4$. This band is comprised between the wavelengths $k=0$ and $k=k_{\rm c}^{[10]}$ where
\begin{flalign}
(k_{\rm c}^{[10]})^2 = \frac{\taue(5/4-\taue)}{\tilde{A}\tilde{D}}.
\end{flalign}

\begin{figure*}[htb]
\includegraphics[width=1.\textwidth]{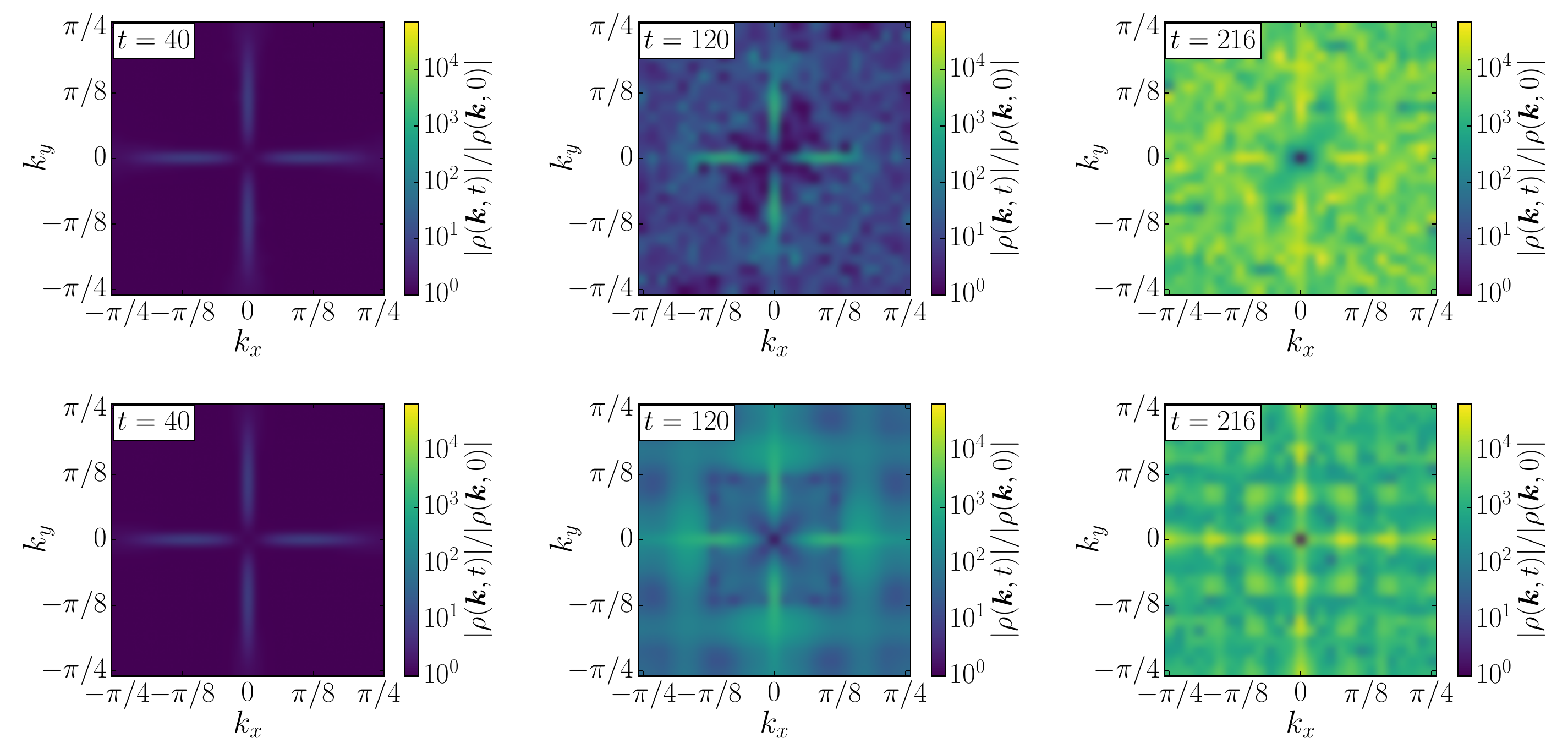}
\caption{Time evolution of Fourier patterns $\rho(\Bk,t)$; top: patterns growing from uncorrelated Gaussian noise (initial condition (i)), bottom: patterns growing from a single localized perturbation (initial condition (ii)); parameters as in Figure \ref{fig:instability}.}
\label{Fig:FFT}
\end{figure*}

Curves $\Lambda(k)$ are shown in Figure \ref{fig:instability}, for fluctuations in the glide directions and along the slip system symmetry axis. The instability occurs for fluctuations aligned with the slip systems, the wavevector of maximum amplification corresponds, for the parameters given in the Figure, to a wavelength of about 12 mean dislocation spacings. Regarding the parameter dependence of the wavelength, the results of Ref. \cite{Wu2017_PRB} carry over: the critical wavelength increases with $A$ and $D$ in approximately linear proportion. 

\subsection{Condition for instability: physical interpretation}

Since instability occurs first in [10] directions, the condition for instability to occur is, in non-dimensional representation, simply given by $\taue < 5/4$ or, in dimensional units, 
\begin{equation}
\taue < (5/4) \alpha \mu b \sqrt{\rho_0}.
\end{equation}
To understand the physical nature of this 
condition, we define the total (scalar) flux of dislocations on slip system $i$ in the homogeneous reference state as
\begin{equation}
j_{i} = \dot{\gamma}_i/b = \sum_s \rho_{i,s} v_{i,s} = \rho_i \frac{b}{B} \left(\taue - \alpha \mu b \sqrt{\sum_{i} \rho_{i}}\right)
\end{equation}
The derivative of the total flux $j_i$ with respect to the slip system dislocation density 
$\rho_{i}$ is then given by
\begin{equation}
\frac{\partial j_{i}}{\partial \rho_{i}} = \frac{b}{B} \left(\taue - \alpha \mu b \sqrt{\sum_{i} \rho_{i}} \left[1 + \frac{\rho_i}{2 \sum_i \rho_i}\right] \right) 
\end{equation}
For the present case where $\rho_i = \rho_0/2$ we thus find that the dislocation density derivative of the total dislocation flux turns negative when $\taue < 5/4 \alpha \mu b \sqrt{\rho_0}$ which is precisely our instability criterion. We are, hence, dealing with a variant of a basic instability that has long been studied in hydrodynamic models of traffic flow, see e.g. \cite{Gerlough1975}. Importantly, no other terms in the evolution equation but the flux term and the friction-like stresses  - which represent the isotropic hardening due to dislocation density accumulation - are needed to observe this instability which is, hence, a quite generic feature of dislocation dynamics.

\section{Numerical analysis}

We have performed a numerical analysis of the evolution equations for two different types of initial conditions, namely (i) a spatially uncorrelated Gaussian white noise of small amplitude and (ii) a localized small perturbation in the origin of the coordinate system. We implement periodic boundary conditions in $x$ and $y$ for the stresses and for the dislocation fluxes on the two slip systems. For the stress evaluation we use a Finite Element framework with periodic displacement boundary conditions. As initial conditions we use $\rho^{\pm}(\Br,t) = \rho_0/2 + \epsilon \delta\rho^{\pm}(\Br,t)$ where $\epsilon \ll 1$ and we consider two types of perturbation $\delta \rho^{\pm}$: (i) a Gaussian white noise of unit amplitude and (2) a localized Gaussian 'blob' of width $l = \rho_0^{-1/2}$ located at the center of the simulation cell. The system is loaded by imposing a constant external stress and keeping it fixed throughout the simulation.  

The time evolution of the Fourier coefficients of the emergent patterns is shown in Figure \ref{Fig:FFT} for both cases. The emergent patterns are dominated by fluctuations with wave-vectors oriented along the symmetry equivalent [01] and [10] lattice directions. From the initial growth rates of the discrete Fourier modes $\rho(k)$ we deduce growth factors defined as $\Lambda(k) = \Delta \ln\rho(k)/\Delta t$. Comparison with the analytical predictions for fluctuations oriented along [10] and [11] lattice directions shows good agreement. The wavelengths of the fully developed patterns match closely (within 20\%) the predictions of linear stability analysis for the wavelength of the mode with maximum amplification. At longer times, satellites appear at multiples of the dominant wavelength and the Fourier spectrum assumes a grid-like pattern, indicating a non-sinusoidal periodic pattern with long-range order. While the initial growth rates of Fourier components are similar for localized and distributed perturbations, the ordering tendency seems to be more pronounced if patterning starts from a single localized perturbation (Figure \ref{Fig:FFT}, bottom). 
\begin{figure*}[p]
\centering
\includegraphics[width=1.\textwidth]{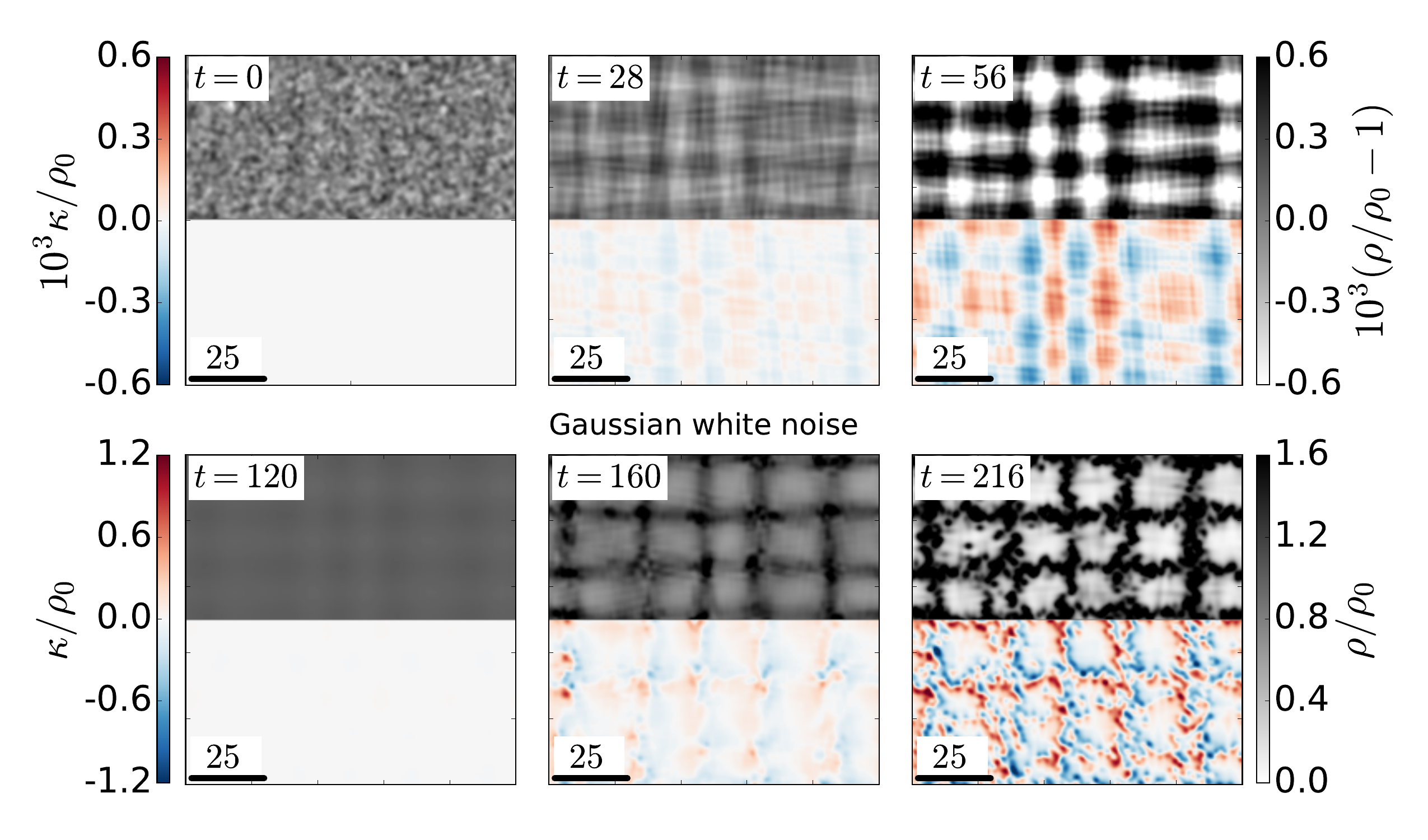}
\includegraphics[width=1.\textwidth]{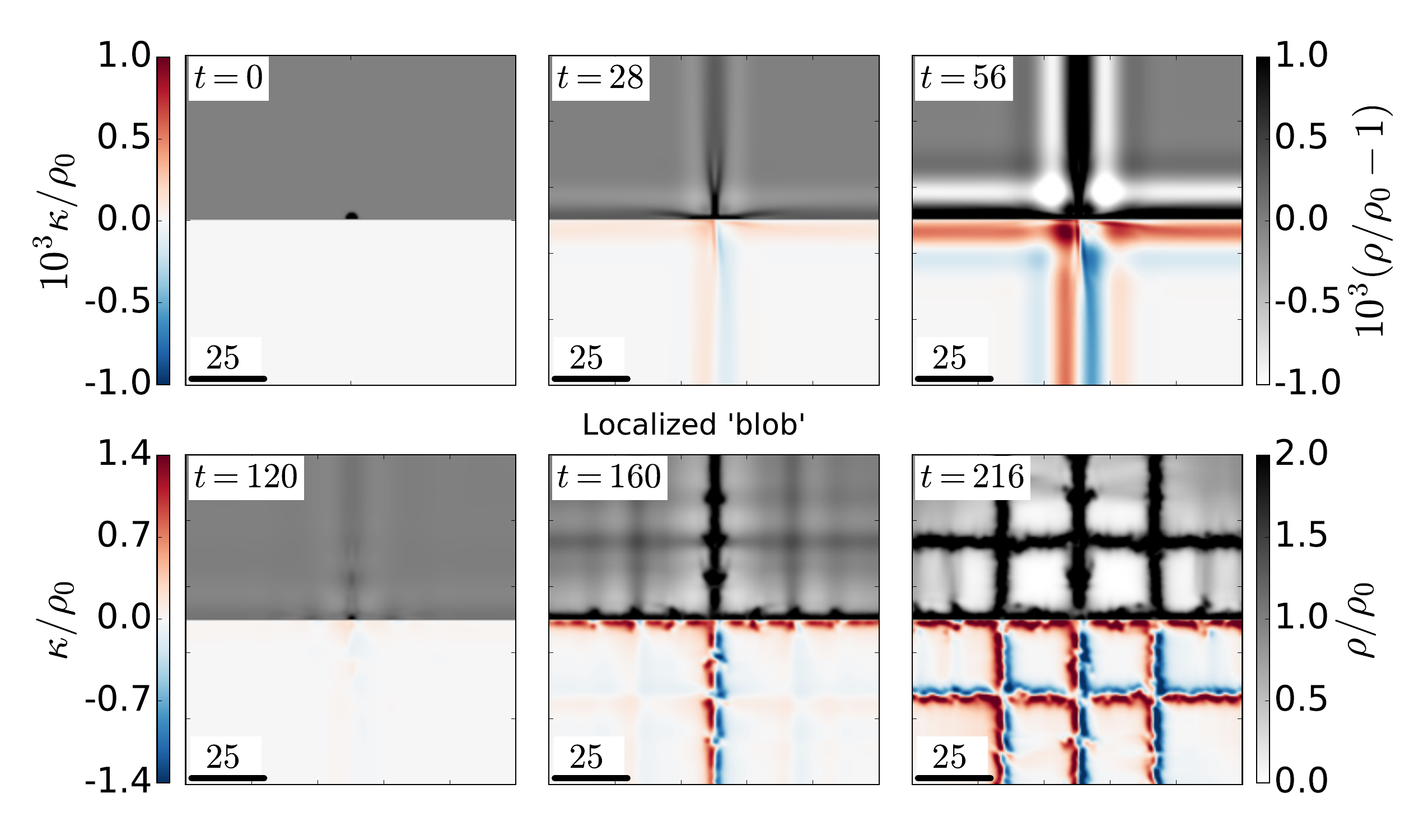}
\caption{Time evolution of spatial patterns $\rho(\Br,t)$ and $\kappa(\Br,t)$  ; top: patterns growing from uncorrelated Gaussian noise (initial condition (i)), bottom: patterns growing from a single localized perturbation (initial condition (ii)); parameters: $D=A=0.2, \taue = 1.1$, these parameters are chosen to match experimental observations shown in Figure \ref{fig:LiF}.}
\label{fig:growth}
\end{figure*}
The mode of growth depends on the initial conditions, see Figure \ref{fig:growth}: in case of a spatially distributed noise the emergent patterns have a crossed stripe-like character. If we use a localized perturbation as initial condition, two perpendicular  walls start growing from the perturbation and then the wall pattern spreads into a grid-like pattern. The characteristic wavelength of the emergent pattern is, however, independent of the growth mode. 
\begin{figure*}[htb]
\centering
\includegraphics[width=1.\textwidth]{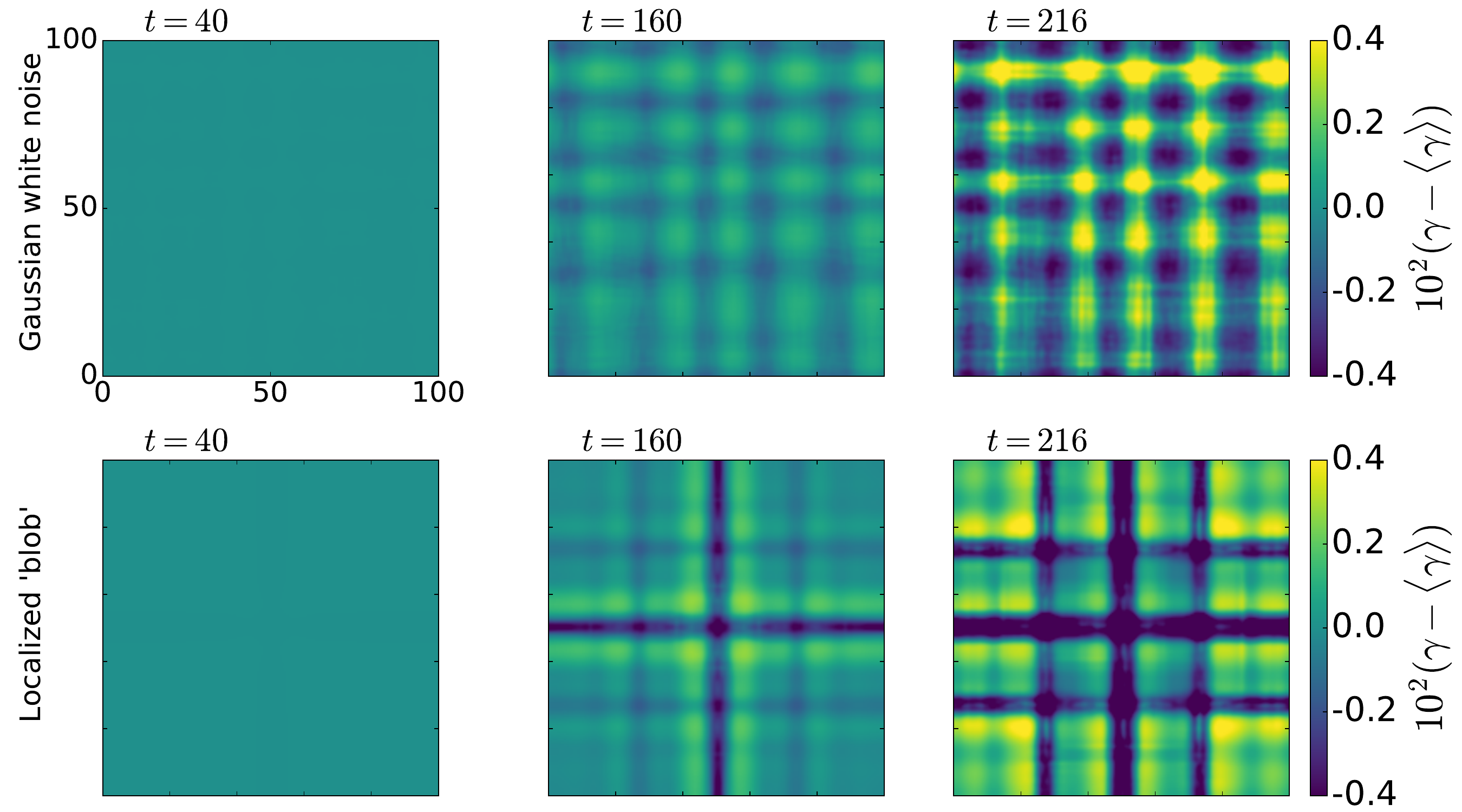}
\caption{Time evolution of the spatial patterns of the local strain fluctuation 
$\gamma(\Br,t) - \langle \gamma \rangle$ top: patterns growing from uncorrelated Gaussian noise (initial condition (i)), bottom: patterns growing from a single localized perturbation (initial condition (ii)); parameters as in Figure \ref{fig:growth}.}
\label{Fig:strain}
\end{figure*}
\begin{figure*}[htb]
\centering
\includegraphics[width=1.\textwidth]{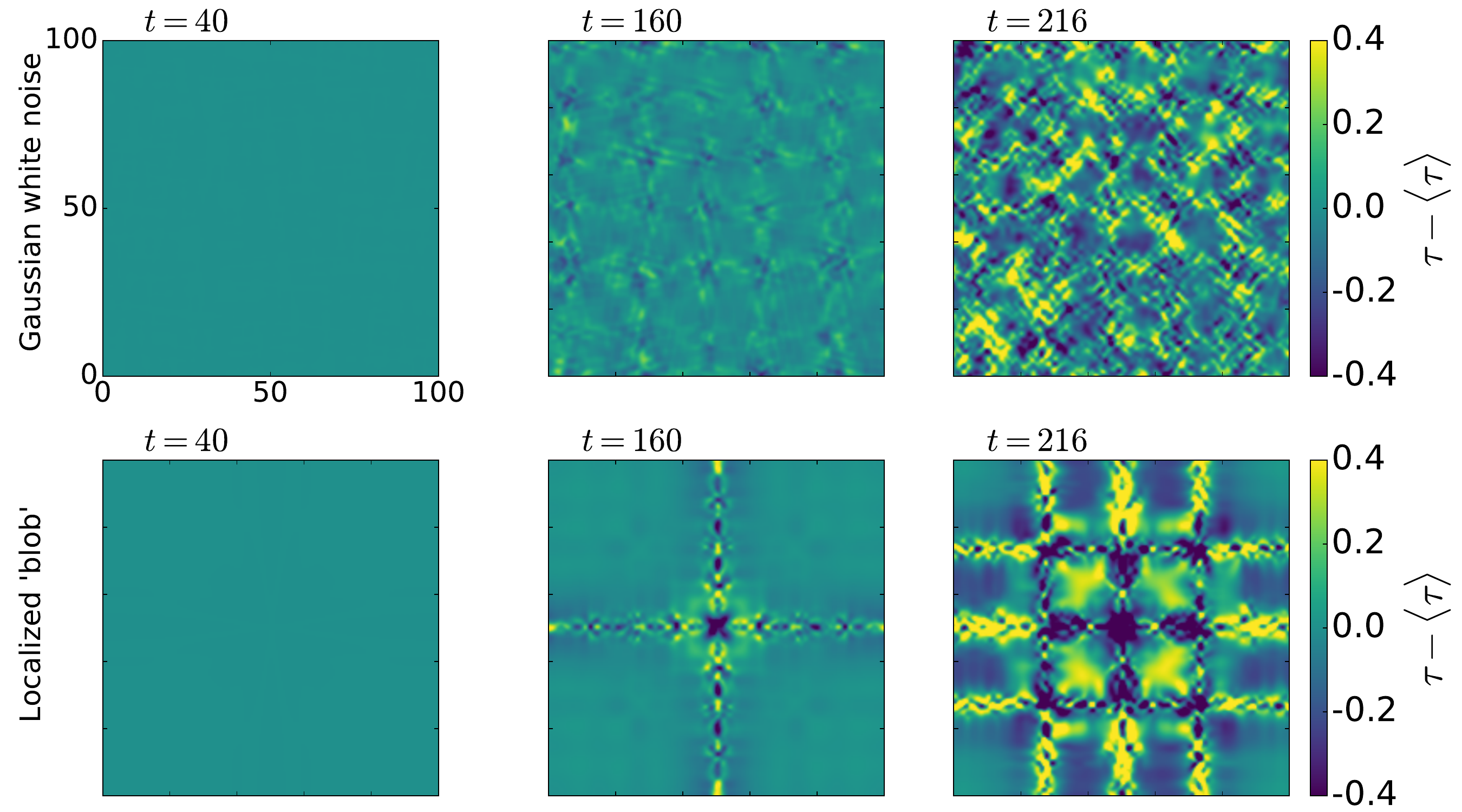}
\caption{Time evolution of the spatial patterns of the long-range internal stress $\tau(\Br,t) - \taue$ and plastic strains $\gamma - \langle \gamma \rangle$; top: patterns growing from uncorrelated Gaussian noise (initial condition (i)), bottom: patterns growing from a single localized perturbation (initial condition (ii)); parameters as in Figure \ref{fig:growth}.}
\label{Fig:stress}
\end{figure*}

An interesting question concerns the applicability (or not) of the well-known composite model to our simulation data. According to the composite model as originally formulated by \cite{Mughrabi1983_AM}, long-range internal stresses associated with slip heterogeneities develop in such a manner as to homogenize deformation. Regions of enhanced dislocation density (cell walls) have a higher local flow stress, accordingly, plastic slip is reduced in these regions. In regions of reduced dislocation density, the flow stress is reduced and slip is enhanced. The compatility requirements between both kinds of regions imply presence of geometrically necessary dislocations which, so the model, create long range internal stresses that offset the flow stress differences.  Ultimately, in quasi-static deformation one expects the local stress to everywhere match the local flow (friction) stress such that deformation can then proceed in a compatible manner: 
\begin{flalign}
\tau(\Br) - \alpha G b \sqrt{\rho(\Br)} = 0 ,\quad \delta \tau = \alpha G b \delta (\sqrt{\rho})
\label{eq:comp}
\end{flalign}. 
Note that this relation is expected to hold independent of the length scale of the pattern: The 'composite' of the original composite model is considered in the spirit of classical composite mechanics which does not know about size effects. The composite model has some important corollaries. For instance, it can be seen immediately that patterning does, in the composite model, always lead to softening (reduction of flow stress) in comparison with the homogeneous reference state: Evaluating the spatial averages $\langle .... \rangle$ and noting that the because of stress equilibrium $\langle \tau(\Br) \rangle = \taue$, we find that in the patterned state because of the triangular inequality $\taue = \alpha G b \langle \sqrt{\rho} \rangle < \tau_{\rm ext,0} = \alpha G b \sqrt{\rho_0}$ where $\rho_0 = \langle \rho \rangle$ is the homogeneous reference density. This finding is supposed to hold independently of the morphology or of the length scale of the heterogeneous patterns (\cite{Zaiser1998_MSEA}). 

Looking at the strain patterns in our simulations we find that they match the expectations: Strain is increased in the cell interiors and decreased in the cell walls. If we look at the internal stress patterns in our simulations, however, a more complex behavior is found. The internal stresses do {\em not} exhibit a strict correlation with the plastic strain, or with the dislocation density, see Figure \ref{Fig:stress}.  

\begin{figure}[htb]
\centering
\includegraphics[width=0.5\textwidth]{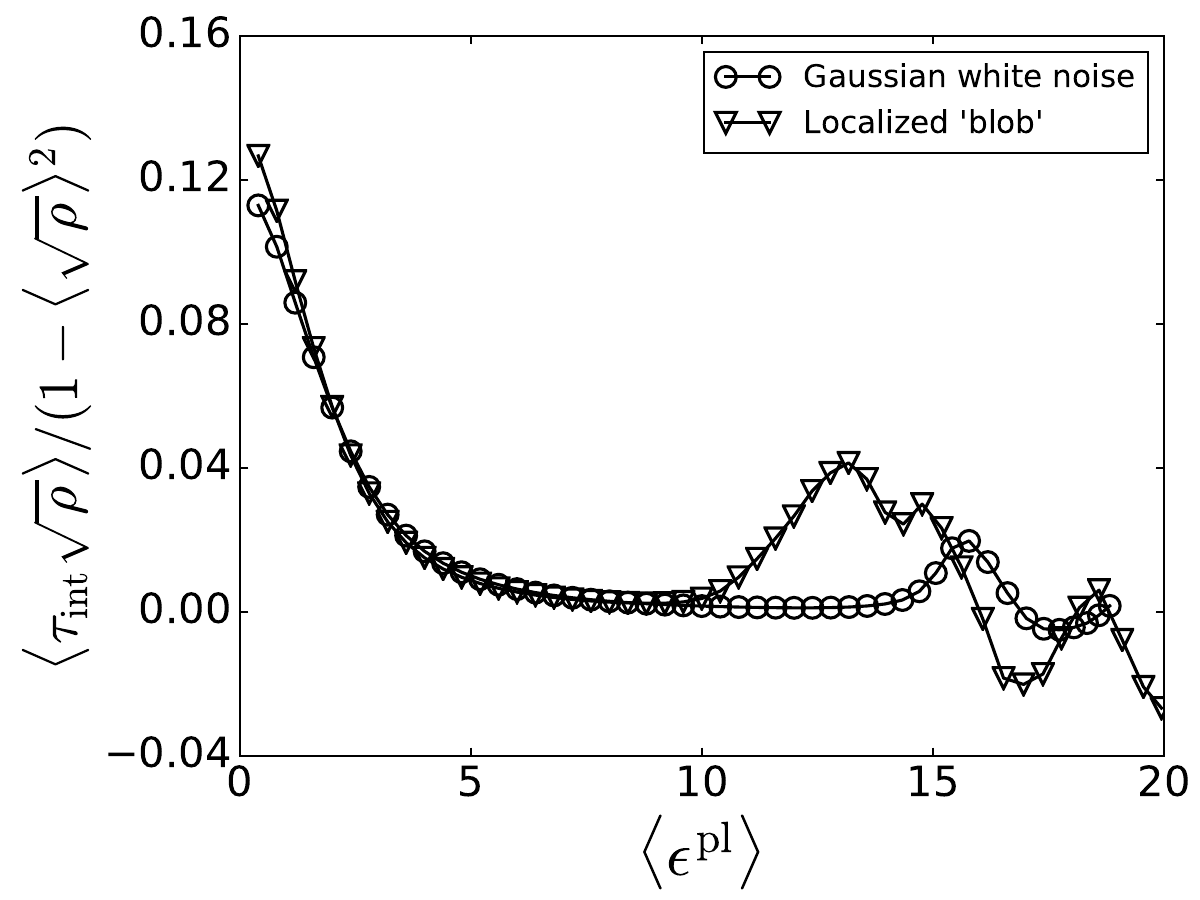}
\caption{Strain evolution of the correlation between internal stress and local flow stress, normalized by the scatter of local flow stresses; parameters as in Figure \ref{fig:instability}.}
\label{Fig:stresscor}
\end{figure}

To quantify the deviation from the composite model, we note that according to the composite model, in non-dimensional variables we expect the local internal stresses and dislocation densities to obey the relation
\begin{flalign}
\frac{\langle (\tau(\Br)-\taue) \sqrt{\rho(\Br)}\rangle}{1 - \langle \sqrt{\rho(\Br)} \rangle^2} = 1
\label{eq:corr}
\end{flalign}
where the angular brackets denote spatial averages. Figure \ref{Fig:stresscor} shows that  a positive correlation which however is significantly below the value expected according to the composite model, exists only during the initial stage of patterning. This correlation actually decreases as patterns are formed and ultimately drops to zero. For patterns emerging from a localized perturbation, there is an additional complication since the correlation oscillates as walls are formed sequentially. Either way, in the fully developed pattern there is no appreciable correlation between local stress and local dislocation density. This raises the intriguing question how the patterns can deform compatibly.

The shortfall is made up by the length scale dependent stress contributions $\tau_i^{\rm b}(\Br)$ and $\tau_i^{\rm d}(\Br)$ which may be considered non-local, strain and dislocation density gradient dependent generalizations of the classical composite model. This points to a limitation of the composite model which assumes an entirely classical composite mechanics framework: If applied to patterns that are heterogeneous on the micrometer scale, where in other composite systems size effects start to become relevant, composite models which neglects non-local stress contributions might systematically under-estimate the flow stress of heterogeneous dislocation arrangements, see also the discussion of strain gradient effects in the composite model in Ref. \cite{Mughrabi2001_MSEA}
\section{Relation to experimental observations}

\begin{figure}[bht]
	\centering
	\hbox{}
	\includegraphics[width=0.4\textwidth]{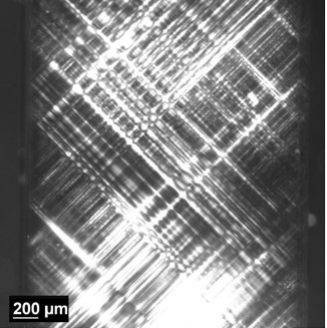}\hfill
	\includegraphics[width=0.46\textwidth]{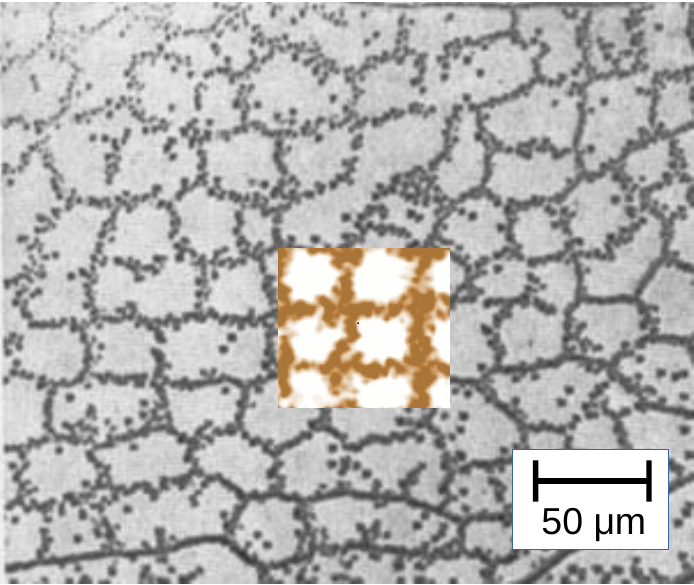}\hfill
		\caption{\label{fig:LiF}
		Cell structures in LiF; top: birefringerence image of the (001) surface of a $(100)$ oriented single crystal
		showing slip activity on the orthogonal $(1\bar{1}0)[110]$ and $(110)[1\bar{1}0]$ slip systems, courtesy of J. Schwerdtfeger;
		bottom: etch pit pattern on a (100) cross section after deformation under a creep load of $\sigma = 5.9$ MPa ($\tau = 2.45$ MPa) to a creep strain of $\epsilon^p = 0.05$    ($\gamma = 0.1$), deformation temperature 773K, averaged dislocation density $\rho = 3 \times 10^{11}$ m$^{-2}$ \cite{Streb1973_PSS}; the insert has been taken from the simulation shown in Figure \ref{fig:growth} and scaled according to the average dislocation density in the experimental image.}
\end{figure}

At first glance a plane-strain slip geometry with two perpendicularly intersecting slip systems as studied in the present idealized model seems unrealistic. However, a quite faithful realization of this situation can be found in early deformation stages of ionic solids with KCl crystal lattice structure. This structure consists of two interlaced fcc sub-lattices containing the K$^+$ and Cl$^-$ ions, respectively. If the crystal is subjected to a uni-axial stress state with the stress axis oriented along the [100] crystal lattice axis, deformation can take place on four symmetrically oriented slip systems which form two conjugate pairs, namely the $(110)[1\bar{1}0]$ and $(1\bar{1}0)[110]$ systems, and the $(101)[10\bar{1}]$ and $(10\bar{1})[101]$ systems. We make the following observations:

\begin{enumerate}
\item
The active slip systems are such that, for tension along a [100] lattice axis aligned with the $x$ axis, the conjugate pairs of active slip systems produce plane strain states in the $xy$ and $xz$ planes, respectively. 
\item
The slip systems in a conjugate pair intersect at right angles. Their mutual interactions are comparatively weak (forming a junction produces, in line tension approximation, no net energy gain). By contrast, there are strong interactions between pairs of slip systems belonging to different conjugate pairs, leading to significant latent hardening.
\item
As a consequence, during the early stage of deformation a symmetry breaking takes place where deformation is taken over by one conjugate pair of slip systems while the second pair becomes inactive \cite{Schwerdtfeger2010_AM}. This situation quite faithfully matches the slip geometry assumed in our simulations.  
\end{enumerate}
Dislocation structures observed in these materials develop heterogeneity already at comparatively small strains, forming cellular patterns as illustrated in Figure \ref{fig:LiF}, right. The wavelength of these structures exceeds the mean dislocation spacing by a factor of about 14. By comparing the patterns with the theoretical results, several important conclusions can be drawn regarding the interpretation of the dislocation density patterns that follow from our model. To this end we remind the reader that all distances are measured in mean dislocation spacings - mds. With a cell size of about 15 mds, we expect on average about 50 dislocation lines threading each cell wall. The walls are essentially dipolar (they carry little net mis-orientation), hence, we expect about 25 positive and an equal number of negative dislocations in a wall. These distribute over a length of 15 mds and a wall thickness of about 5 mds, hence, the density is in the wall increased by a factor about 3, as consistent with the simulations. Owing to the imbalance of fluxes during wall formation, dipoles form preferentially in such a manner that positive and negative dislocations gather on the opposite sides of the wall. The width of dipoles can be estimated by noting that the dislocations forming a dipole stem from independent sources, hence, it will be of the order of (1/5) mds which, with a typical dislocation density of $\rho = 3 \times 10^{11}$ m$^{-2}$, translates into a spacing of the dislocations in the dipoles of the order of about 0.35 $\mu$m, well above the atomic spacing. Hence, annihilation of dislocations is not expected to be a relevant process here.

The walls are formed by the mutual trapping of dislocations into dipole-like configurations (friction stress). They are stabilized by two effects that mutually compensate each other: On the one hand, excess of dislocations of positive sign pushes against the wall from one side ('pile up stress') , on the other hand, the dislocations within a dipole push each other back ('diffusion stress'). As a consequence we see a wall consisting of polarized dipoles, with positive and negative dislocations accumulating on opposite sides of the wall. The width of the walls, the corresponding width of the cells and the dislocation spacings are all in good agreement with the experimental observations. This can be seen in Figure \ref{fig:LiF}, right, where a piece of the simulated dislocation density pattern could, after re-scaling to the dislocation spacing in the experiment, be seamlessly pasted into the experimental image.  

\begin{figure}[tb]
\centering
\hbox{}
\includegraphics[width=0.6\textwidth]{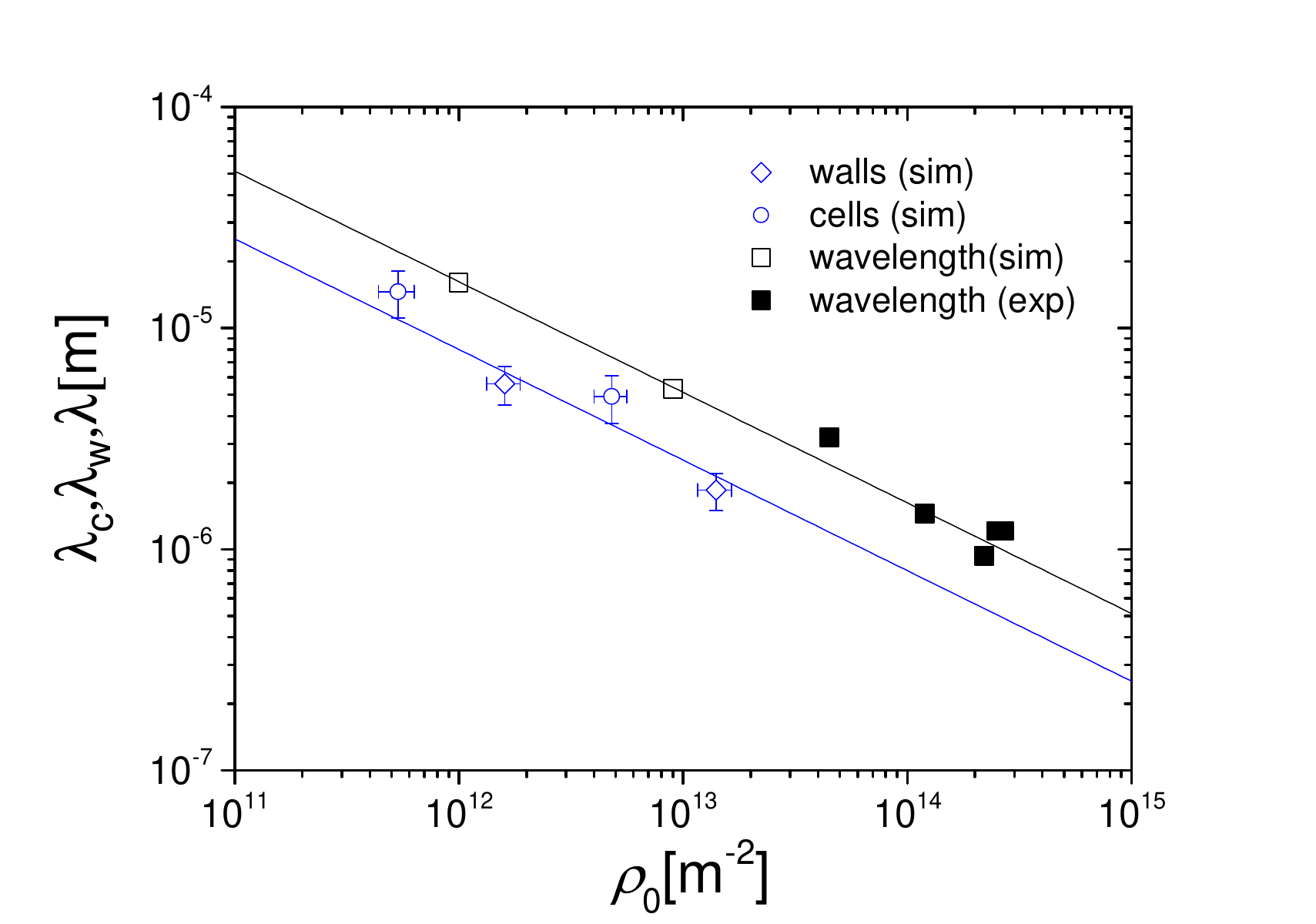}\hfill
\caption{\label{fig:scaling}
Length scales vs dislocation densities in simulated cell structures; open circles: cell interiors, open diamonds: cell walls, the error bars indicate the standard deviation of data obtained from 10 interiors/walls determined from intercept method as explained in text; open squares: overall pattern wavelength vs average dislocation density; full squares: experimental pattern wavelength vs average dislocation density data \cite{Oudriss2016_IJP}.}
\end{figure}

We also investigate whether our patterns match the similitude principle in the strong form proposed in Ref. \cite{Oudriss2016_IJP}. To this end we study one-dimensional density profiles taken along the slip directions and define, for a given profile, the wall dislocation density $\rho^{\rm w}_i$ of wall $i$ as the dislocation density at the corresponding density maximum and the channel dislocation density $\rho^{\rm c}_i$ as the dislocation density in the corresponding density minimum. Left and right wall boundaries $x_i^{\rm l}$ and $x_i^{\rm r}$ are defined as the locations where the dislocation density takes the respective values $(\rho^{\rm w}_i-\rho^{\rm c}_i)/2$ and $(\rho^{\rm w}_i-\rho^{\rm c}_{i+1})/2$. The width of wall $i$ is then evaluated as $\lambda_{i}^{\rm w} = x_i^{\rm r} - x_i^{\rm l}$ and the width of channel $i$ as $\lambda_{i}^{\rm c} = x_i^{\rm l} - x_{i-1}^{\rm r}$. Figure \ref{fig:scaling} shows lengths $\lambda^{\rm c,w}$ as well as pattern wave-lengths $\lambda$against the corresponding densities $\rho^{\rm c,w}$ for different values of the average density $\rho_0$. As can be seen, the data are well represented by a common fit function $\lambda^{\rm c,w} = C \sqrt{\rho}^{\rm c,w}$ with $C \approx 6$, in good agreement with the findings of Oudriss et al \cite{Oudriss2016_IJP}. Also the overall relationship between pattern wavelength $\lambda = \lambda^{\rm c}+\lambda^{\rm w}$ and total dislocation density $\rho_0$ matches well experimental data \cite{Oudriss2016_IJP} (full data points in Figure \ref{fig:scaling}). We thus conclude that our model is consistent with the strong similitude principle as observed by Oudriss et al. \cite{Oudriss2016_IJP}.
\section{Discussion and Conclusions}

We have presented a very simple model of dislocation cell structure formation in a 2D setting with two perpendicularly intersecting slip systems. Despite its simplicity, the model can be considered a elementary representation of dislocation processes in a real system, namely a crystal with KCl lattice structure deformed uni-axially along a cube axis. We find formation of cellular dislocation patterns with a cell size of the order of about 10 mean dislocation spacings. The patterns obey the similitude principle: their wavelength is proportional to the dislocation spacing and inversely proportional to the stress at which they form. The simplicity of the 2D model, which can not account for dislocation multiplication, does not allow us to consider strain hardening. However, if we impose a higher overall dislocation density $\rho_0$, then deformation requires an accordingly higher stress that scales in proportion with $\sqrt{\rho_0}$, and similitude is maintained. 

It is instructive to discuss our findings in relation to commonly held viewpoints on dislocation patterns: (i) It is an often expressed viewpoint (see e.g. \cite{Madec2002_SM, Xia2015_MSMSE} that cross slip is essential for dislocation cell structure formation. However, it is easy to see that in KCL structures, as in our simulations, this mechanism is irrelevant since there is only one (110) slip plane for each [110] slip vector, hence, there are no cross-slip planes. Nevertheless, formation of cellular dislocation patterns is observed regularly in these structures and our simulations - where cross slip is excluded by construction of the model - provide an excellent match to the observed cellular patterns. We therefore conclude that cross slip is, in the end, incidental to dislocation patterning. (ii) The composite model predicts that a patterned dislocation arrangement deforms at a stress that is strictly below the stress needed for deforming a homogeneous reference arrangement. This assumption is predicated upon a classical treatment of internal stresses that does not allow for strain gradient dependent effects. Even within the classical continuum mechanics framework, it is clear that dislocation patterns or strain patterns of general morphology in general produce internal stress patterns that do not directly match the strain/dislocation patterns as required by the composite model, compare our Figures \ref{Fig:stress} and \ref{Fig:strain}. In fact, for the present slip geometry a match between stress and dislocation patterns would be possible only if the dislocation patterns would form with a [11] orientation which they do not. Deformation compatibility must therefore be ensured by other means that cannot be described by standard continuum mechanics. Such effects are also needed to understand pattern wavelength selection. In our model these effects are provided by the gradient dependent stress contributions $\tau^{\rm b}$ and $\tau^{\rm d}$, in other models a similar role is played by curvature related terms  \cite{Sandfeld2015_MSMSE}.  (iii) The only essential requirement for patterning in our model is that, for a given stress, the local dislocation flux is a decreasing function of local dislocation density. Many models of work hardening fulfill this requirement for a wide range of deformation parameters. We therefore conclude that, if dislocation density evolution is described by appropriate transport equations, patterning is an expected feature of dislocation dynamics. Our investigation can be easily generalized to a wide range of stress-velocity laws in order to provide guiding principles that allow to decide under which deformation conditions heterogeneous patterns may form. It thus provides an important complement to microstructure-based plasticity models as proposed e.g. by \cite{Castelluccio2017_IJP} which investigate the impact of self-organization of dislocations into mesoscale structures on the macroscale deformation behavior under complex loading paths. 

Regarding the conditions for patterning, we may note that, in standard tensile testing, the axial strain rate rather than the external stress is imposed. It is therefore instructive to re-phrase our patterning criterion in terms of an imposed strain rate in the homogeneously flowing reference state. For the deformation geometry at hand, the axial strain rate in that state (Schmid factor 1/2, 2 active slip systems) is simply $\dot{\epsilon}_0 = \dot{\gamma_0}$. The instability
condition, Eq. (35), can then be written as 
\begin{equation}
\dot{\epsilon} = \frac{\rho_0 b}{2B} \left(\taue - \alpha \mu b \sqrt{\sum_{i} \rho_{i}}\right) \le 
\rho_0^{3} \mu b^3 \frac{\alpha}{8B} 
\end{equation}
We re-write this in terms of a non-dimensional parameter combining dislocation density, strain rate, and material constants:
\begin{equation}
P = \left(\frac{\mu b^3}{4 B}\right)^{2/3} \frac{\rho_0}{\dot{\epsilon}^{2/3}} \ge P_{\rm c} = \left(\frac{2}{\alpha}\right)^{2/3} 
\end{equation}
This critical parameter $P_{\rm c}$ separates a regime where the flow stress decreases with increasing dislocation density (no patterning) from 
a regime where the flow stress increases with dislocation density (patterning). Remarkably, a recent study by Fan et. al. \cite{Fan2020_NC} demonstrates that the same parameter also controls the shape of the dislocation velocity distribution and the magnitude of dislocation velocity fluctuations, separating a regime of large fluctuations (large $P$) from a regime of small fluctuations (small $P$). In conjunction with the present findings we see that dislocation controlled plasticity exhibits two regimes: a quasi-laminar regime with small fluctuations and homogeneous flow at high strain rateslow dislocation densities (small $P$) and a quasi-turbulent regime with large fluctuations, unstable dislocation flow, and dislocation patterning at low strain rates/high dislocation densities (large $P$).  
\section*{Declarations}
\subsection*{Competing interests}
  The authors declare that they have no competing interests.
\subsection*{Author's contributions}
R.W. implemented the simulation model and performed the simulations. M.Z. performed the stability analysis. Both authors jointly wrote the 
manuscript.
\subsection*{Funding}
M.Z. acknowledges funding by DFG under Grants no. 1 Za 171/7-1 and 1 Za 171/13-1. 
\subsection*{Acknowledgements}
Not applicable
\subsection*{Availability of data and material}
Not applicable

\bibliographystyle{bmc-mathphys} 
\bibliography{literature}   
\newpage

\end{document}